# Diversity against adversity: How adaptive immunity evolves potent antibodies

Muyoung Heo, Konstantin B. Zeldovich, Eugene I. Shakhnovich

Department of Chemistry and Chemical Biology

Harvard University, 12 Oxford Street, Cambridge, MA 02138

**Abstract**

How does immune system evolve functional proteins – potent antibodies - in such a short time? We address this question using a microscopic, protein-level, sequence-based model of humoral immune response with explicitly defined interactions between Immunoglobulins, host and pathogen proteins. Potent Immunoglobulins are discovered in this model via clonal selection and affinity maturation. Possible outcomes of an infection (extinction of cells, survival with complete elimination of viruses, or persistent infection) crucially depend on mutation rates of viral and Immunoglobulin proteins. The model predicts that there is an optimal Somatic Hypermutation (SHM) rate close to experimentally observed $10^{-3}$ per nucleotide per replication. Further, we developed an analytical theory which explains the physical reason for an optimal SHM program as a compromise between deleterious effects of random mutations on nascent maturing Immunoglobulins (adversity) and the need to generate diverse pool of mutated antibodies from which highly potent ones can be drawn (diversity). The theory explains such effects as dependence of B cell fate on affinity for an incoming antigen, ceiling in affinity of mature antibodies, Germinal Center sizes and maturation times. The theory reveals the molecular factors which determine the efficiency of affinity maturation, providing insight into variability of immune response to cytopathic (direct response by germline antibodies) and poorly cytopathic viruses (crucial role of SHM in response). These results demonstrate the feasibility and promise of microscopic sequence-based models of immune system, where population dynamics of evolving Immunoglobulins is explicitly tied to their molecular properties.

**Introduction**

Adaptive immunity is one of the marvels of Biology and one of its greatest mysteries. Organisms have to respond to novel antigens which they have not seen before and this response should be specific in order to avoid attacking its own proteins (autoimmunity). This task represents a clear example of adaptation, which occurs in a shortest possible time frame of days to weeks, while normal evolutionary processes of adaptation take anywhere from years to millions of years (Elena and Lenski 2003). The adaptive humoral immunity is based on the ability of Immunoglobulins to bind antigen ligands and to quickly evolve protein-protein (in case of protein-based antigen) or more generally protein-ligand interactions. Immunoglobulin-based adaptive immunity represents a rapid and effective search in protein sequence space under a number of constraints. Understanding basic principles of immune response, besides its obvious importance for medicine, is also important for fundamental understanding of Darwinian evolution and adaptation.

The modern view on sequence evolution in immune response stems from Burnet hypothesis that selection of B-cells expressing potent Immunoglobulins occurs in two stages (Burnet 1959). At the first stage (clonal selection) a small fraction of naïve antibodies that are able to bind the antigen with moderate affinity are found in a diverse germline pool. Subsequently, these cells are activated to proliferate and they accumulate mutations which increase its affinity to antigen (affinity maturation). Affinity maturation has been indeed discovered (Eisen and Siskind 1964) and subsequent experimental studies provided important insights into molecular mechanisms of adaptive immunity consistent with Burnet's model (Jacob et al. 1991; Jacob and Kelsoe 1992; Liu et al. 1992; Shih et al. 2002). Experiments found that some specificity for an antigen indeed exists in germline pool of Immunoglobulins before it is encountered (Eisen and Siskind 1964; Jacob and Kelsoe 1992; Rajewsky 1996; Paus et al. 2006). Recent real time imaging experiments clarified many mechanistic aspects of affinity maturation taking place in Germinal Centers (GC) (Allen et al. 2007).

These important developments notwithstanding, it is difficult to grasp fundamental physical principles of adaptive immunity from the molecular detail picture emerging in experiments. It may be not apparent which evolutionary ''solutions'' for mechanism of immunity are robust and which ones represent the result of particular random evolutionary ''choices''. It is as if one were to try to grasp the principles of thermodynamics by dissecting modern internal combustion engines.

Here we take a complementary bottom-up approach by studying adaptive humoral immunity within a model which is realistic enough to contain sequence-based description of stability of proteins and their interactions yet is sufficiently coarse-grained to allow study of immune response in an ab initio simulations and analytically. Our main goal is to uncover the minimal requirements and physical principles on which a successful adaptive immune response can be built and to investigate whether existing immune systems are based on these physical principles.

Recently we developed a model which combines microscopic, sequence-based, description of dynamics of genes and stability of encoded proteins with cell population dynamics governed by exact relationship between genotype and phenotype (Zeldovich et al. 2007b). This model was applied in (Zeldovich et al. 2007b) to study early evolution of Protein Universe. Here we extend this approach to study adaptive humoral immunity. We explicitly determine stability and interactions between viral and host defense (Immunoglobulin-like, Ig) proteins from their sequences. On a cellular level our model accounts for replication of host cells, viruses and B cells (see below) and mutations of their sequences. We build the model bottom up based on simplest ''common-sense'' mechanistic assumptions and sequence-based molecular description of relevant proteins and their interactions. (see Figure 1) Specifically we make the following four mechanistic microscopic assumptions. *First,* we assume that viruses invade host cells and cause their death by lysis when the number of viral particles replicating in a cell exceeds the lysis threshold. *Second*, we assume that stronger binding between Ig proteins and viral antigens inhibits replication of viruses (i.e. decreases effective viral replication rate). *Third,,* to account for B-cell activation mechanism, we assume that the rate of B-cell division increases upon stronger binding between viral antigens and Ig proteins. *Finally* we assume that B-cell activation is impaired if Ig proteins bind cell's own proteins, roughly modeling the effect of protection against autoimmunity, via clonal exclusion of helper T-cells. For simplicity, we model the complex processes of B-cell activation and antibody production, differentiation and excretion, via the dependence of dynamics of production of Ig proteins on their physical interactions with the antigen (virus) and host cell proteins. Therefore, in the model, B-cells, antibodies, and Ig are merged into a single entity (an Ig protein) whose population dynamics is separate from that of host cells and viruses. Although all three terms are used in the text as appropriate, we stress that in the model, they all refer to the same object, Ig protein. The technical details of the model are given in the Methods section and in the Supporting Information.

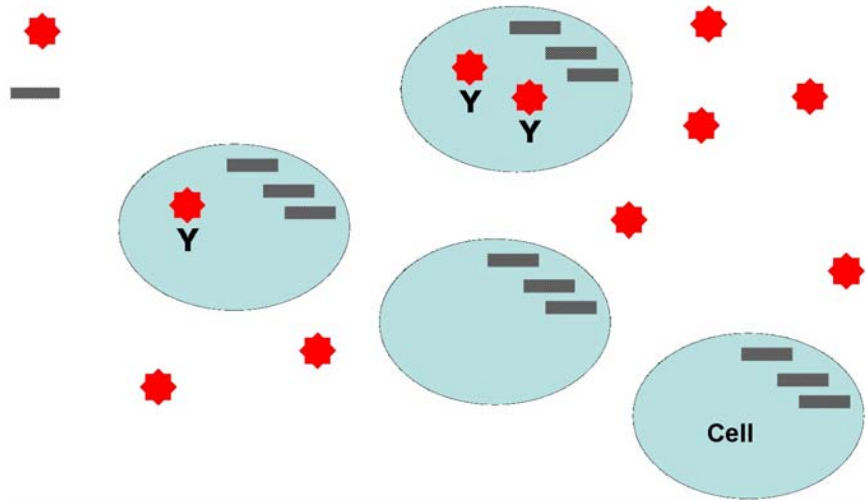

**Figure 1.** *Schematic representation of the model. Viruses (red stars) may penetrate cells and reproduce there, eventually causing lysis. Antibodies (Y-shaped symbols) bind to viruses and slow down their replication. The replication rate of the antibodies increases with the strength of their interaction with viruses (B-cell activation) and decreases with the strength of their interaction with host cell proteins (autoimmunity).*

We show that an effective adaptive immune response very similar to real one develops in this microscopic model. Most importantly, we demonstrate that our model of adaptive dynamic immune response captures many complexities of real immune systems and helps to explain a variety of recent observations concerning immune response to pathogens in several experimental models.

Dynamics of the system establishes a qualitative outcome of an infection in this model: healing (survival of host cells with complete elimination of viruses), extinction of host cells, or development of a persistent infection, where the cells and viruses coexist and the fraction of infected cells remains constant with time. Qualitatively, one can expect that for a very slowly mutating virus, strongly virus-binding Ig proteins will evolve, suppress virus growth, and eliminate the virus population. On the other hand, the speed at which such immunity evolves is fundamentally limited by the mutation rate of the Ig protein, and by the speed at which the advantageous sequences spread through the population. Therefore, if the mutation rate of viruses is high enough, their sequence change can outpace the evolution of cellular response, resulting in a robust lethal or persistent infection.

Below, we present the results of simulations and analysis for this ab initio, microscopic model of immune response, and show that physics-based, sequence-level model can provide crucial insights, on all scales, into the development and dynamical regimes of immune response. Further we present a detailed analytical theory that provides quantitative description of evolution and selection of potent Immunoglobulin sequences emerging in response to an Antigen presentation.

**Results and discussion**

Extensive simulations of the model over a broad range of parameters (B-cell and virus replication rates, and mutation rates of Ig and viruses) showed that the outcome of an individual simulation run falls into one of the three categories: healing, H, with complete extinction of the virus; extinction of the host, E, (and the ensuing extinction of the virus); and persistent infection, P, where viruses and cells coexist with neither species winning the competition. The probability of each of the three outcomes depends on the parameters of the model, most notably on mutation rates of Ig and viral genes.

Figures 2A and 2B represent a phase diagram of the model, where the color corresponds to the probability of healing (Figure 2A) and persistent infection (Figure 2B) at the given mutation rates of virus $m_v$ and of Ig protein $m_{Ig}$. For low-to-moderate virus mutation rates, the most notable result is the existence of the optimal Ig mutation rate, which maximizes the probability of healing for the given $m_v$ (Figure 2A and Figure 7). Qualitatively, for a very low Ig mutation rate, strongly binding Ig sequences can not evolve before viruses spread in the population and cause lysis, eliminating the host cells. On the other hand, if the Ig mutation rate is too high, favorable Ig sequences mutate away from strong binding before their carrier antibodies proliferate and eliminate the virus. An analytical estimate of the optimal mutation rate of the Ig sequences is presented below. At higher viral mutation rates, the typical outcome of a simulation run is either extinction or persistent infection (Figure 2B), as the antibodies can not keep up with the fast-mutating viral population, which thus effectively evades immunity.

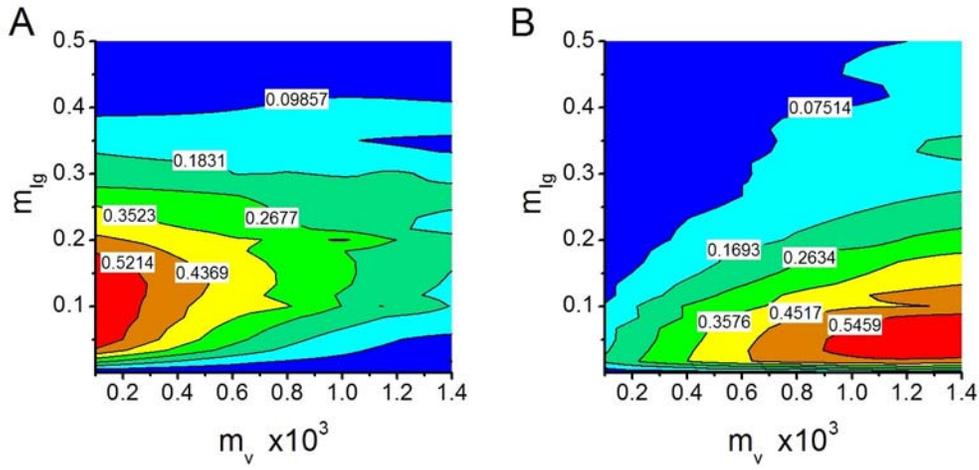

**Figure 2.** *Emergence of immunity. (A) Phase diagram showing the probability of healing of host cells in 200 runs as function of the mutation rates of Ig and viral proteins, in mutations per protein per time step. (B) Phase diagram showing the probability of persistent infection. The labels on the contour lines are the probabilities of healing in (A) or persistent infection in (B).*

Figures 3A and 3B present the population dynamics of cells and viruses, respectively, for representative simulation runs corresponding to healing (black curves), persistent infection (blue curves), and extinction (red curves) cases. One can see that in all cases, infection is followed by an incubation period during which the host cell population remains constant, and the virus population increases until lysis threshold is achieved in the majority of cells. Ensuing lysis causes an abrupt drop in the cell population (Figure 3A). The duration of the incubation period depends on the virus replication rate. At the same time, B-cells producing antibodies which strongly bind to viruses, acquire selective advantage, as their replication rate is higher, while virus replication is hindered by strongly bound Ig.

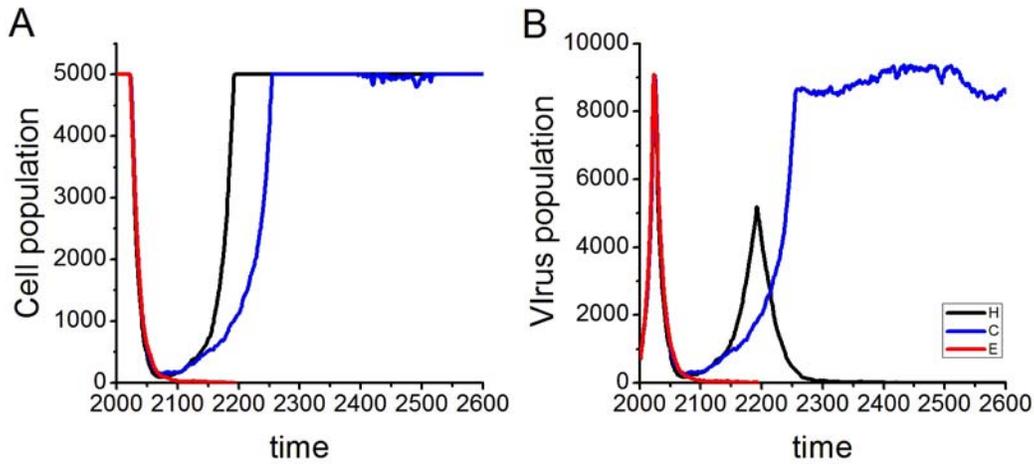

**Figure 3.** *Population dynamics of the model (A) Population of cells as a function of time in representative simulation runs resulting in healing (H) at $m_v=0.0003$ and $m_{Ig}=0.08$, extinction of host cells (E) at $m_v=0.0005$ and $m_{Ig}=0.04$ or persistent infection (P) at $m_v=0.0008$ and $m_{Ig}=0.001$. Infection occurs at $t=2001$. (B) Population of viruses in the same simulation runs. Red lines in **A-B** correspond to extinction outcome, blue lines to persistent infection and black lines to healing.*

As a result the average interaction strength between Ig and virus proteins $P_{int}$ (see Methods) increases with time (Fig.4). In a particular run (infection event), the outcome (in a certain range of viral, cell and B-cell replication rates and viral and Ig mutation rates) largely depends on random events during the competition of Ig and viral sequences, affecting binding between Ig and viruses, their population dynamics and ultimately the fate of the host cells.

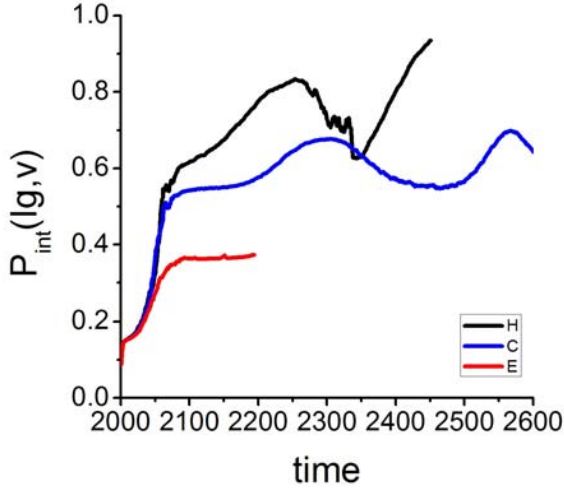

**Figure 4** *Averaged (over all Ig proteins and viruses) strength of interaction $P_{int}$ between viral and Ig proteins shows that strongly binding Ig's gain selective advantage, rapidly increasing $P_{int}$ as the cells try to overcome infection.*

To understand the behavior of the model, let us note that healing occurs mainly via dilution of viruses within the host cells, when the cell replication rate $b^{cell}$ exceeds that of viruses $b^v$. Using Eq.(21) of Methods which defines dependence of viral replication rate on antigen-Ig interaction strength we get:

$$b^{cell} > b^v = b_0^v P_{nat}(v)[1 - P_{int}(Ig,v)]$$

or (1)

$$P_{int}(Ig,v) > 1 - \left(b^{cell}/b_0^v P_{nat}(v)\right) \sim 0.8$$

for the replication rates used in the model (see Methods). Thus, healing outcome may occur only if sufficiently strong binding between Ig proteins and viruses, $P_{int} > 0.8$ evolves. Further growth of affinity beyond $P_{int} = 0.8$ does not confer selective advantage and is therefore limited. Conversely, lysis occurs when the population of viruses within an infected cell $p_v$ reaches the lysis threshold $L$ before cell division, i.e. over times on the order of $1/b^{cell}$. A similar calculation shows that such outcome is possible for $P_{int}(Ig,v) < 1 - \left((L-1)b^{cell}/b_0^v P_{nat}(v)\right) \sim 0.4$. Indeed, as seen in Figure 4, in the extinction case the interaction strength $P_{int}$ never exceeded 0.35 (no strong-binding Ig had evolved), and was significantly higher in cases of persistent infection and healing, with healing observed when Ig-antigen interactions are strongest. Lower and upper thresholds on affinity in B-cell activation were also observed in experiment (Batista and Neuberger 1998).

**Germline immune response to evolving viruses – the advantage of being a moderate parasite.**

An important ecological implication of the above considerations is a non-monotonic dependence of effective virus fitness on strength of its interaction with germline Ig. Indeed, viruses which are prone to strong binding to Ig (high $P_{int}$) replicate slowly and are removed from population via dilution as their carrier cells divide faster than the viruses within the cells replicate. On the other hand, very weakly binding viruses are so efficient at evading immune defense that they quickly destroy their carrier cells through lysis, leading to the extinction of the populations of *both* host cells and viruses. This can be seen most strikingly in simulations with a fixed antibody repertoire, representing germline Igs i.e. $m_{Ig}=0$. We find that viruses evolve a certain intermediate interaction strength, steering away both from the removal by immune response and from prematurely killing every host cell. Figure 5 shows histograms of interaction strength $P_{int}$ for $m_{Ig}=0$ for two values of virus mutation rate, $m_v=10^{-4}$ (Figure 5A) and $m_v=10^{-1}$ (Figure 5B). For the lower value of virus mutation rate, the viruses are so efficient at evading and destroying the host (under non-evolving immunity conditions) that their population dies out (together with the host) while retaining extreme virulence (low $P_{int}$). On the other hand, the population of rapidly mutating viruses ($m_v=10^{-1}$) is diverse enough to evolve towards maximum fitness, which results in a persistent infection, with gradual *upward* evolution of the $P_{int}$ distribution towards a stationary, moderately virulent state. Interestingly, the interplay between molecular and ecological effects in this model results in a host-pathogen coexistence where pathogens require some level of host defense (immune response) for their survival. A very similar behavior, where the primary defense is provided by naïve serum germline antibodies, was observed with highly cytopathic viruses such as VSV (Roost et al. 1995; Hangartner et al. 2006).

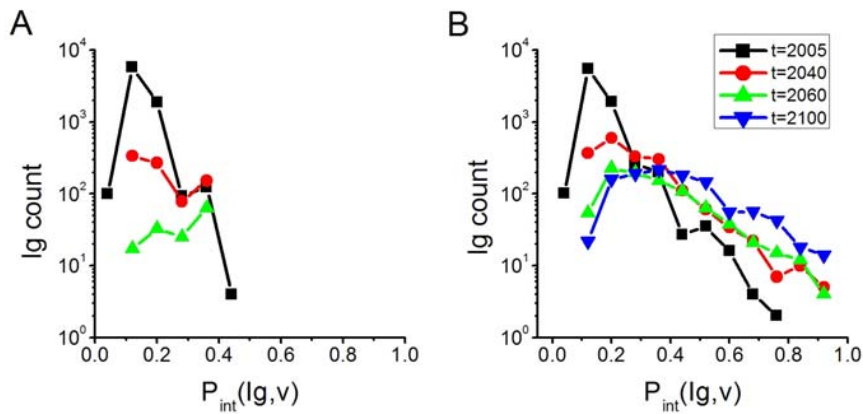

**Figure 5.** *Coevolution of viruses and host cells with germline Igs only, i.e., $m_{Ig}=0$. (A) A slowly mutating virus remains extremely virulent, $P_{int}<0.4$, and kills the host and the virus population. (B) At a higher virus mutation rate, development of moderately strong interactions between viruses and Ig leads to a persistent infection, as the Ig can not eliminate the virus, and the virus, in turn, is too weak to kill the host.*

**Clonal Selection and Affinity maturation**

Now we consider higher Ig mutation rates reflective of the phenomenon of somatic hypermutaion (SHM) in GC (Jacob et al. 1991; Kleinstein et al. 2003). Our main finding here is that at higher Ig mutation rates, selection of specific, strongly binding Ig proteins proceeds in our model via clonal selection with subsequent affinity maturation. In Figure 6A we present the histogram of the Ig-to-virus binding strength $P_{int}$ distributions at four time slices. Immediately after infection, $t=2005$ (black curve), most of the Ig sequences exhibit a marginal binding to the viral antigen with $P_{int} \sim 0.10$, typical of two random proteins in this model. However, a small fraction of B-cells expresses Ig molecules with relatively strong binding, $P_{int} > 0.35$ (tail of the distribution, shaded area). Subsequently, such cells enjoy a selective advantage – they are activated to divide faster (see Eq. (3) in Model and Methods), and their proportion in the population increases ($t=2030$, red curve). This is similar to the natural clonal selection process (Jacob et al. 1991; Mehr et al. 2004) Furthermore, after (moderately) binding sequences have been discovered, they undergo mutations with subsequent selection, corresponding to affinity maturation increasing the binding strength $P_{int}$ to its final value of ~0.7 ($t>2050$, green and blue curves). To confirm that strongly binding Ig's are normally direct descendants of the original moderately binding sequences selected at the first – clonal selection – stage of the immune response, we marked initial antibodies with $P_{int}>0.35$ as ''red'', and postulated that their progeny retains the ''red'' color. In Figure 6B, we plotted the fraction of the ''red'' antibodies in the population. One can see that the fraction of the ''red'' antibodies increases with time, and approaches unity as the system recovers from the infection (cf. Figure 3). Therefore, virus removal is mostly accomplished by direct descendants of the germline B-cells expressing Ig sequences which accidentally had a certain binding affinity to the incoming antigen and have been subsequently amplified and optimized via a two-stage process of clonal selection and affinity maturation. Our simulations suggest that the primary role of clonal selection is to amplify B-cells with moderately binding sequences from a diverse germline pool. As a next step, cells carrying clonally selected Ig genes undergo improvement by specific mutations dramatically increasing the binding strength. Evidently, such mutations can be efficiently discovered only during mutagenesis in a pool of moderate binders, rather than by pure chance among pseudorandom sequences.

Next, we study the distribution of sequences of evolving Ig proteins. In Figure 6C, we present the measure of sequence diversity - sequence entropy - of evolved Ig proteins (black curve) and Ig-to-virus binding strength $P_{int}$ (red curve) as a function of time for a healing case. The plot shows that discovery of strongly binding Ig proteins, manifested in a sharp increase in $P_{int}$, is accompanied by an initial drop in sequence entropy of Ig proteins – reflecting strong selection of mature Ig molecules resulting in an almost monoclonal character of emerging mature antibody pool. As the infection is being eliminated, both population and sequence diversity of Ig proteins increase. Notably, an increase in sequence entropy does not lead to a decreased interaction strength $P_{int}$ suggestive of availability of large number of mutations which do not affect binding affinity. This is reflective of significant plasticity in sequence space. Our finding that population of mature strongly binding Ig is highly diverse, yet it is monoclonal in its ancestry is in accord with classical experimental observations that individual GCs produce mature monoclonal B cells (Jacob et al. 1991).

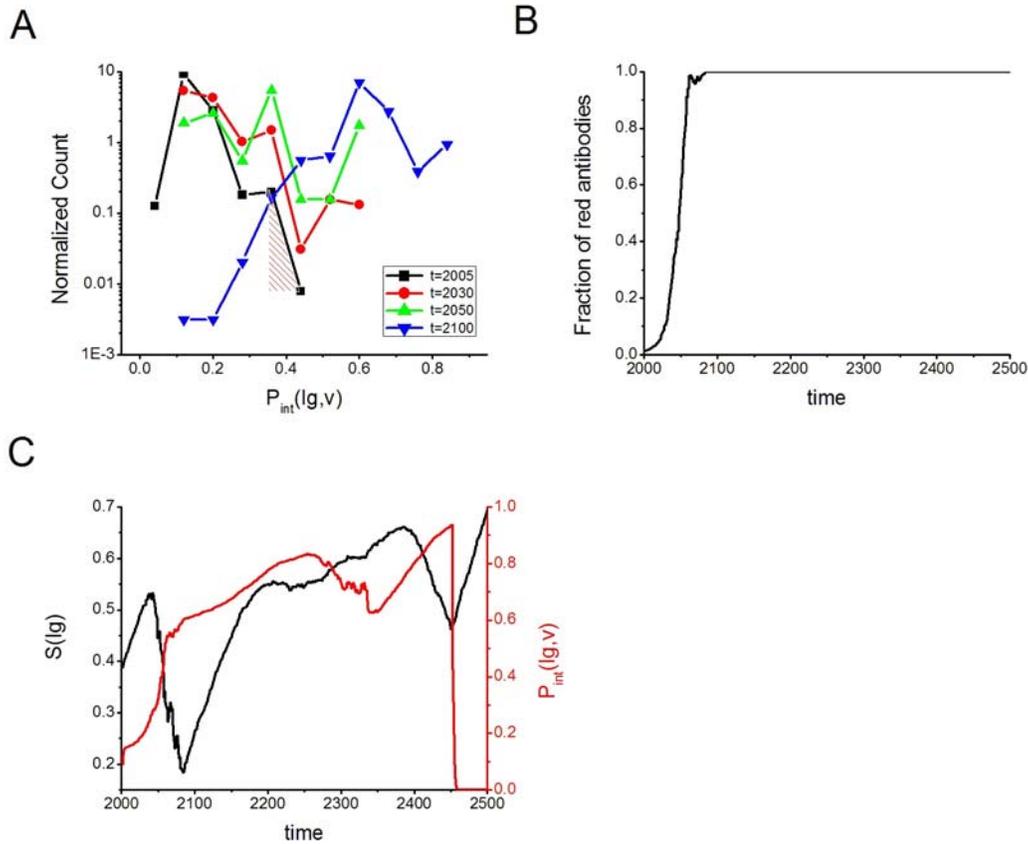

**Figure 6.** *Clonal selection and affinity maturation. (A) Histogram of the binding strength $P_{int}$ in the ensemble of Ig molecules at various times. From t=2001 to t=2030, clonal selection increases the fraction of Ig's strongly binding the virus: the tail of the distribution becomes more prominent, while the maximum is unmoved. As time goes on, mutations in previously selected Ig genes further increase $P_{int}$ and shift the distribution to the right until monoclonal population of strongly binding Ig molecules emerges. (B) We marked the antibodies with $P_{int}>0.35$ at t=2001 (shaded area in (A)) as ''red'' and followed the fraction of their progeny in subsequent populations. As the infection is being eliminated, most of the Ig's are descendants of the few strongly binding sequences amplified via clonal selection. (C) Sequence entropy of the evolving Ig proteins (black line) and their binding strength $P_{int}$ as a function of time (red). Development of immunity is initially accompanied by a rapid increase of $P_{int}$ and a decrease of the sequence entropy of Ig proteins, confirming the appearance of a monoclonal Ig population at the initial stage of infection. However subsequently, the pool of Immunoglobulins gets more diverse resembling of a polyclonal population.*

**Theory of affinity maturation and optimal rate of SHM.**

Our simulations suggest the existence of optimal rate of SHM at which immune response is most efficient (see Figure 2A and (Celada and Seiden 1996; Pierre et al. 1997)). Qualitatively, this is due to the fact that too low mutation rate of Ig proteins will prevent affinity maturation while too high mutation rate could result in prevalence of predominantly deleterious mutations which destroy even moderate binding of Ig to antigens found at the clonal selection stage.

A more quantitative analysis is based on the MacLennan model of affinity maturation in GCs (MacLennan 1994; Allen et al. 2007). Early proliferation of B cells in GC occurs prior to formation of Dark Zone (DZ) and Light Zone (LZ). SHM accompanied by exponential growth of B cells takes place in DZ. B cells in DZ (centroblasts) downregulate expression of Igs (Allen et al. 2007) and therefore are not subject to selection at this stage. Expansion of centroblasts in DZ creates a diverse repertoire of mutated sequences of $V_{H,L}$ regions of Igs. Selection takes place most likely in the LZ where high concentration of Ag is present after which B-Cells expressing Ig's with sufficiently improved affinity are positively selected and the ones showing low affinity Igs are removed via apoptotic pathway (Berek et al. 1991; Jacob et al. 1991; Batista and Neuberger 1998; Allen et al. 2007). While the molecular details of this positive selection have not been fully established we based our theoretical analysis on the following dynamic postulates:

1) B cells having some initial affinity to the antigen (free energy of binding) $G_0$ clonally expand to become centroblasts in the DZ (Berek et al. 1991).

2) In the DZ centroblasts enter cell cycle where they proliferate and undergo stochastic mutation program over t generations with mutation rate m (per $V_{H,L}$ genes per generation). Each cell acquires a random number of mutations *N*, varying from cell to cell due to intrinsically stochastic nature of SHM. As Ig genes accumulate mutations in the process of SHM, the total change in binding free energy for Ag is additive with respect to mutations in CDR of this gene.

In order to understand conceptually why this picture implies the existence of an optimal SHM program consider first an (oversimplified) case when CDRs of all $V_{H,L}$ genes acquire an identical number of mutations $mt \gg 1$. The total change in binding free energy between Ag and Ig is additive with respect to all mutations. Additivity of free energy effects of mutations implies that probability density for the total free energy change $\Delta G$ over complete SHM program follows a Gaussian distribution according to the Central Limit Theorem:

$$P(\Delta G) = \frac{1}{\sqrt{2\pi mt\sigma^2}} e^{-\frac{(\Delta G - mth)^2}{2mt\sigma^2}} \qquad (2)$$

where

$$h = \int_{-\infty}^{\infty} p(\Delta G)\Delta G d\Delta G; \quad \sigma^2 = \int_{-\infty}^{\infty} p(\Delta G)\Delta G^2 d\Delta G - h^2$$

are average and dispersion of change of binding free energy upon a single mutation; $h > 0$ implies that mutations in the binding interface *on average* weaken binding, which is indeed the case in experiment (Chen et al. 1999). Now we assume that upon exit from DZ into LZ only those B-cells are clonally expanded whose Ig binding free energy to Ag reaches or falls below the threshold value $G_c$. In this case the probability of successful affinity maturation (i.e. fraction of B cells which differentiate into Ig producing plasma cells or memory cells) is apparently:

$$P_{AM} = \int_{-\infty}^{G_c-G_0} P(\Delta G) d\Delta G \approx erfc\left(-(G_c-G_0)\right) \approx -\frac{1}{\sqrt{\pi}} \frac{\sqrt{mt}\sigma}{G_c-G_0-mth} e^{-\frac{(G_c-G_0-mth)^2}{2mt\sigma^2}} \quad (3)$$

where we used an asymptotic expression for the complementary error function assuming that $(G_c - G_0 - mth)/\sqrt{mt}\sigma \gg 1$. When $h>0$ this expression reaches a maximum at

$$(mt)_{opt} = -\frac{G_c - G_0}{h} \quad (4)$$

The reason that random mutations in CDR tend to be destabilizing in average is because B cells enter the DZ of GC after initial clonal selection with some initial potency to bind Ig with binding free energy $G_0 < 0$. In that case random mutations on the binding surface of Igs produced by these initial centroblasts would have a tendency to weaken this initial binding affinity. In other words an estimate of $h$ for a binding surface

$$h \geq -\alpha G_0 \quad (5)$$

where $\alpha \sim 1$ describes how mutation in a single amino acid changes binding free energy in a potent interface. The probability of successful affinity maturation at the optimal SHM program when $mt = (mt)_{opt}$ is then

$$P_{AM}^{opt} = e^{-\frac{2(G_c-G_0)h}{\sigma^2}} \quad (6)$$

This quantity increases when $G_0 \to G_c$, which is natural because the range of necessary affinity increase shrinks in this case.

This simple example illustrates the physical reason for an optimal mutation rate and/or duration of SHM (SHM program). Indeed, as random mutations accumulate they in average weaken binding. However by virtue of additivity of energetic effects of individual mutations, they create a more diverse pool of B-cells (i.e. dispersion of binding free energies between B-cells after N mutations grows as $\sigma\sqrt{N}$) and this diversity increases the chance that there appear B-cells which produce sufficiently strong binding Igs which then can further expand and differentiate into plasma or memory cells.. Clearly a compromise between average mutational destabilization of already binding surfaces and diversity can be achieved only at some optimal total number of mutations in the SHM program (see Fig.7A for a graphic illustration).

The above illustrative considerations pertain to an oversimplified case when all centroblasts experience the same number of mutations during their SHM program. In reality the number of mutations acquired in a centroblast during SHM program is itself random, the total number of mutations is additive over generations and therefore the probability that a centroblast acquires $N$ mutations can be well approximated by a Gaussian:

$$I_N \approx \frac{1}{\sqrt{2\pi mt}} e^{-\frac{(N-mt)^2}{2mt}} \quad (7)$$

(the relation between average and standard deviation in Eq.(7) follows from the fact that the number of mutations at each round of replications follows a Poisson distribution)

The fraction of B-cells which underwent successful AM is then given by:

$$P_{AM} = \int_{-\infty}^{G_c - G_0} d(\Delta G) \sum_{N=0}^{\infty} I_N P_N(\Delta G) \quad (8)$$

Using the asymptotic expression for the complementary error function, Eq.3, and Eq.4 for $(mt)_{opt}$ we get:

$$P_{AM} \approx \sum_{N=0}^{\infty} \frac{e^{-\frac{h^2}{2\sigma^2} \frac{\left((mt)_{opt} + N\right)^2}{N} - \frac{(N-mt)^2}{2mt}}}{2\left((mt)_{opt} + N\right)\sqrt{\frac{h^2}{2\sigma^2 N}} \sqrt{2\pi mt}} \quad (9)$$

We approximate the sum in Eqs. (8,9) by its maximum term in N corresponding to most probable number of accepted mutations $N = \tilde{N}$ in selected B-cells. Denoting the average fraction of accepted mutations $x = \frac{\tilde{N}}{mt}$ we get in the same asymptotic approximation the equation that determines the largest term (in $N$) in series of Eq.(9)

$$\frac{h^2}{2\sigma^2} \left(\frac{(mt)_{opt}}{mt}\right)^2 \frac{1}{x^2} + 1 - \frac{h^2}{2\sigma^2} = x \quad (10)$$

This equation determines the most probable total number of mutations in *matured* B cells which were selected to become plasma or memory cells.

Equation (10) can be solved numerically. It is also important to note that for the optimal SHM program $mt = (mt)_{opt}$ it has a solution $x=1$ which corresponds to highest probability that B cells successfully mature.

Substituting (10) into (9) we get for the probability that a B cell matures:

$$P_{AM} = \frac{e^{-\frac{h^2}{2\sigma^2} \frac{mt\left(x+\frac{(mt)_{opt}}{mt}\right)^2}{x} - \frac{mt(x-1)^2}{2}}}{2\left((mt)_{opt} + xmt\right)\sqrt{\frac{\pi h^2}{\sigma^2 x}}} \quad (11).$$

Where $x$ is determined from solution of Eq.(10).

Antibodies which matured to reach sufficient affinity (i.e. binding free energy stronger than $G_c$) clonally expand upon their exit from DZ (O'Connor et al. 2006). We deem AM to be successful if at least a single mature B-cell (out of $M$ total generated in GC) exists in the centroblast population which may be sufficient for subsequent clonal expansion. Using Eq.11, we finally obtain the probability that AM is successful:

$$P_{GC} = 1 - e^{-MP_{AM}} \quad (12)$$

This result imposes the condition on the number of B cells in GC which have to be generated to assure a successful immune response:

$$M \geq \frac{1}{P_{AM}} \quad (13)$$

These theoretical results can be compared with simulations. The healing outcome in simulations occurs when strongly binding Igs with $P_{int} > 0.8$ evolve and clonally expand (see Eq.1 and Fig.4), which implies that

$$P_{healing} = CP_{GC} \quad (14)$$

We plot the probability of healing in simulations $P_{healing}$ as a function of Ig mutation rate (see Fig.7B). In order to compare theory with simulations we need to know the following parameters: $C, M, t, \frac{h^2}{2\sigma^2}, m_{opt}$. We can estimate most of them directly from simulations. $m_{opt}$ is determined as a mutation rate $m$ at which healing probability is maximum, $t \approx 50$ can be estimated as time when diversity of Ig binding significantly increases indicative of appearance of mature Ig molecules (see Fig.6) and $M=8000$ is population of Ig cells in simulation. That leaves us with only two adjustable parameters, C and $\frac{h^2}{2\sigma^2}$. Comparison of theory with simulations is shown in Fig.7B.

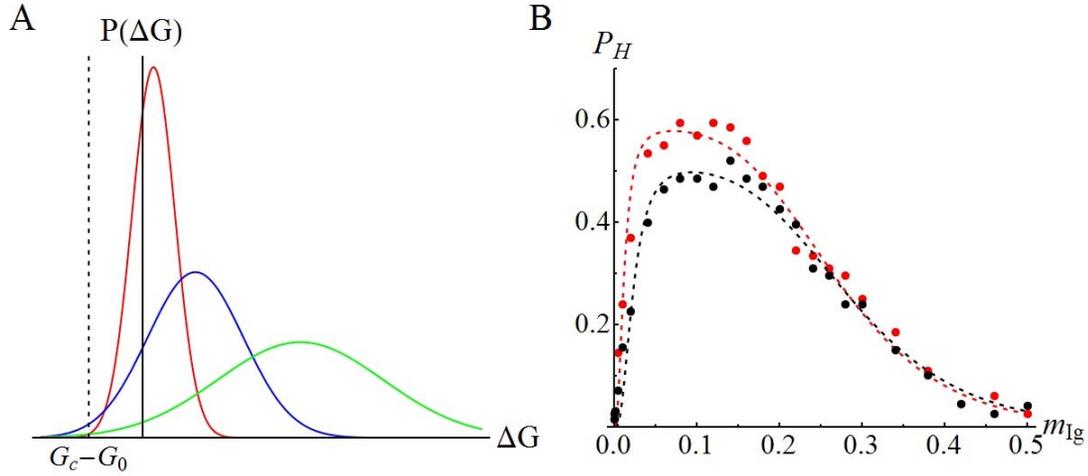

**Figure 7**. *The existence of an optimal SHM rate of Ig. **(A)** A cartoon plot provides a conceptual explanation of why an optimal mutation rate exists. Three solid lines in the plot represent the distribution of binding energy variation after characteristic period ($t_{opt}$) of affinity maturation at low (red), optimal (blue), and high (green) mutation rate. The more mutations are accumulated, the more rightward shift occurs because the majority of random mutations are deleterious to the affinity of Ig antigen interactions. The probability of the affinity maturation is the area under the distribution curve to the left of the dashed line, marking the threshold of the binding free energy for affinity maturation. At low mutation rate (red line) too few mutations have occurred, therefore the width of the distribution is too narrow to allow affinity maturation. The probability of affinity maturation is still low at high mutation rate, because the peak of the distribution is too far away to cross the dashed line, despite of the wide variation of the distribution – too many deleterious mutations have occurred. At the optimal mutation rate (blue line), the probability is maximized by the moderate width of the distribution and shift of the peak.*
***(B)** Probability of healing as function of Ig mutation rate for the viruses whose mutation rate is $m_v$=0.0001 (red) and 0.0003 (black). Dots are actual simulation data; dashed lines are obtained by nonlinear regression fit of the analytical expression Eq. (14) with $P_{AM}$ from Eqs.(11,12) using Mathematica 6.0. The approximate values of $m_{opt}$, $\frac{h^2}{2\sigma^2}$ and M are obtained from simulation data as 0.1 (0.12), 0.5 (0.3), and 8000 (8000) for $m_v$=0.0001 ($m_v$=0.0003) respectively, and two fitting parameters d and $t_{opt}$ were determined by fitting Eq. (12) to the probability profile from the simulation. The fitted parameters are d=0.60 (d=0.53) and t=31.68 (t=43.64). The optimal activation time of 43.64 at $m_v$=0.0003 is very close to the optimal activation time in simulations (~50, see Fig. 6).*

The optimal SHM program $(mt)_{opt}$ depends on the initial affinity of the clonally selected antigen, i.e. $G_0$: The optimal mutation rate and/or duration of SHM program decreases when Ags of higher affinity to germline Igs are presented as can be discerned from Eq.4. Do organisms adjust their SHM program in response to a presented Ag to keep it close to optimal or there is a preset SHM program, i.e. $mt$ is constant regardless of which antigen is presented? The experimental literature on that subject is somewhat controversial. Authors of Ref (Shih et al. 2002) argue that there is a fixed SHM program while the data of Noelle and coworkers may imply that the SHM program (i.e. the number of mutations) adjusts to variation in initial Ag affinity (O'Connor et al. 2006). Both papers compare the numbers of mutations observed in CDRs with mutations in fragments which are presumably not under selection for affinity (intron in (Shih et al. 2002) and framework (FR) region in (O'Connor et al. 2006)). While in the former case the number of mutations was the same for high- and low- affinity antigens it was dramatically different in the latter case providing some support to the view that the mutation rate and/or duration of the SHM program may adjust to specific antigens. Below we briefly discuss both scenarios.

*a) Case of adjustable SHM program.*

In this case x=1, $mt \approx mt_{opt}$ and Eq.11 can be further simplified to give:

$$P_{AM}^{adj} = \frac{\exp\left\{-4\frac{h^2}{2\sigma^2}mt_{opt}\right\}}{2\frac{h}{\sigma}(mt)_{opt}\sqrt{\pi}} \quad (15)$$

The duration of SHM program decreases and the proportion of matured B-cells grows exponentially upon increase of initial antigen affinity, i.e. when $G_0 \to G_c$. This is probably the case when memory B cells exist for an incoming antigen. Experiment shows that the number of mutations acquired in mouse model $V_{H,L}$ $mt = (mt)_{opt} \approx 15$ (Shih et al. 2002). It is more difficult to estimate the parameter $\frac{h^2}{2\sigma^2}$ for interactions between Ig and Ag. However a similar (in spirit) estimate from experimental data was made in (Zeldovich et al. 2007a) for mutations affecting protein stability. Using these data as a rough approximation we get an estimate $\frac{h^2}{2\sigma^2} \approx \frac{1}{6}$ giving finally for the optimal size of a GC $M \approx 10^4$ B cells which is roughly consistent with reality (MacLennan 1994). (This is a rough estimate, because (Zeldovich et al. 2007a) dealt with energetic effect of mutations on protein stability while this quantity concerns the effect of mutations on free energy of interactions between Ag and Ig)

*b) Case of preset SHM program* (Shih et al. 2002)

In this case $mt = const$ regardless of which antigen is presented. However the actual number of mutations accepted in matured B cells differs from the most probable number of mutations *mt* due to post-SHM selection. The strength of selection can be quantified by parameter $x = \frac{N}{mt}$ introduced above. This quantity can be obtained by solving Eq.10. (see Fig.8A). The initial affinity of germline Igs to antigen $G_0$ determines– the stronger initial affinity $G_0$ is, the lower is $(mt)_{opt}$ (see Eq.(4)). The result is shown in Fig.8B. It is in agreement with experiment: presentation of higher affinity Ag (i.e. lower $mt_{opt}$) results in a smaller number of accepted mutations in selected B cells (Shih et al. 2002). However the observed decrease of the number of accepted mutations in $V_{H,L}$ genes when higher affinity antibodies are presented is also consistent with the possibility of adjustable SHM program because in this case the actual SHM program *mt* varies to match $(mt)_{opt}$ when various Ags are presented.

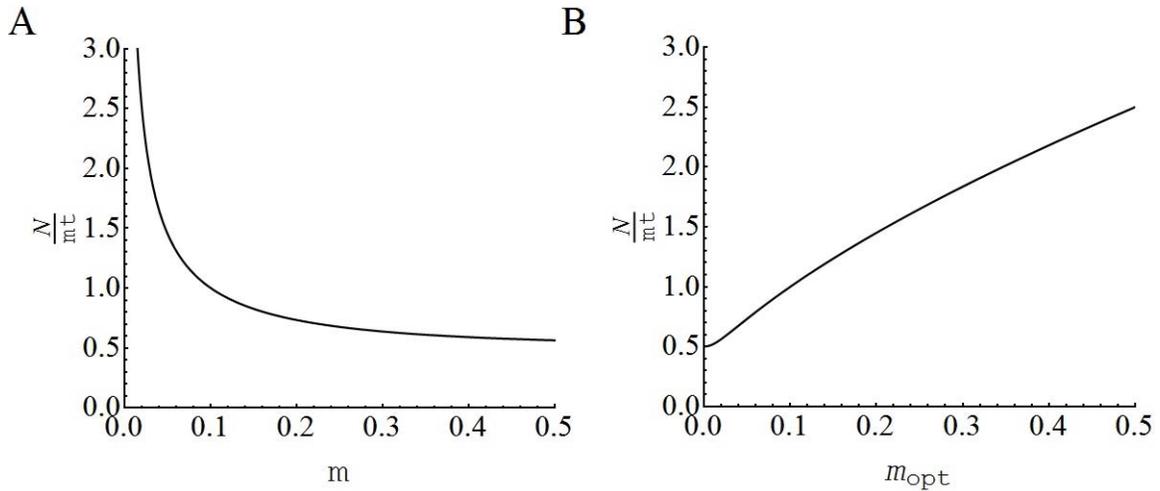

**Figure 8** *Selection in a SHM program. (A) The number of mutations in selected mature cells as a function of SHM rate. This number is normalized by average number of mutations, so that this figure highlights the role of selection. (B). The strength of selection depends on the affinity of the incoming Ag. Lower initial affinity (i.e. greater $G_0$), corresponds to a more extensive optimal SHM program (greater $(mt)_{opt}$) and increased number of accepted mutations (at a given fixed average number of attempted mutations mt). A similar behavior is observed in experiment (Shih et al. 2002). The results of (A) and (B) are obtained from solution of Eq.(10)*

**Why and when is SHM needed?**
At first glance the two-stage process where first Igs of moderate affinity $G_0$ are selected and then they undergo additional round of mutations may seem redundant. Why not select (admittedly few) strongly binding Ags from a huge germline pool and clonally expand those without additional round of mutations? In other words an alternative scenario could be a formation of the same GCs, activation of moderate affinity centroblasts there so that each centroblast clonally expands into M copies but without SHM, i.e. all M centroblasts in any GC express the same Ig in this scenario (while centroblasts of different GCs express different Igs) . In order to determine which scenario is advantageous we compare the number of mature centroblasts (i.e. which express Ig with binding free energy to antigen stronger than $G_c$) under each scenario.

In the first scenario (''one-shot'' selection without SHM) the number of mature centroblasts is:

$$N^{1s}(G_c) = \frac{MN_g}{\sqrt{2\pi\Sigma^2}} \int_{-\infty}^{G_c} e^{-\frac{(G-G_{av})^2}{2\Sigma^2}} dG \approx -\frac{MN_g}{\sqrt{\pi}} \frac{\sqrt{\Sigma}}{G_c - G_{av}} e^{-\frac{(G_c-G_{av})^2}{2\Sigma^2}} \quad (16)$$

where we assumed that binding free energies between Ag and germline pool Igs are Gaussian distributed (Deeds et al. 2006; Lukatsky et al. 2007), $N_g$ is the total number of B cells in germline pool. $G_{av}$ is average free energy of interactions between germline Igs and Ag and $\Sigma$ is the variance of free energy between Ag and Ig over all Igs in the germline pool.

In the two stage process involving SHM GC are established first in a ''one-shot'' selection process i.e. by recruiting germline B cells expressing Igs with affinity exceeding a certain threshold value $G_0 > G_c$ and then centroblasts clonally expand to M cells in each GC, undergoing a SHM program.

The number of mature centroblasts obtained under the 2-stage SHM scenario is apparently:

$$N_{AM}^{2s}(G_c) = N_{GC} M P_{AM}^{adj}(G_c - G_0) \approx N_g M \frac{\Sigma \sigma}{2\sqrt{\pi}(G_0 - G_{av})(G_c - G_0)} e^{-\frac{(G_0 - G_{av})^2}{2\Sigma^2}} e^{2\frac{(G_c - G_0)h}{\sigma^2}}$$

(17)

where $N_{GC} = N^{1s}(G_0)$ is the number of GCs formed and we assumed that SHM program is optimal –best case scenario for 2-stage selection . The probability $P_{AM}^{adj}(G_c - G_0)$ is given by Eq.(15) and its dependence on $G_c - G_0$ is through $(mt)_{opt}$ according to Eqs.(4,5). The system has flexibility in the choice of the affinity threshold $G_0$ at which to activate SHM program. An optimum with respect to the choice of $G_0$ is achieved at:

$$G_0^{opt} = G_{av} - 2h\frac{\Sigma^2}{\sigma^2} \quad (18)$$

Substituting (18) into (17) we obtain maximal possible efficiency of the 2-stage program of development of potent B cells via clonal selection and affinity maturation:

$$\left(N_{AM}^{2s}\right)_{max} = -\frac{\sigma^3 N_g M}{4\sqrt{\pi}h\Sigma\left(G_c - G_{av} + 2h\frac{\Sigma^2}{\sigma^2}\right)} e^{\frac{2h^2}{\sigma^2}\left[\frac{G_c - G_{av}}{h} + \frac{\Sigma^2}{\sigma^2}\right]} \quad (19)$$

The two-stage program is advantageous over one-shot selection when $G_0^{opt} > G_c$, i.e. when

$$G_c < G_{av} - 2h\frac{\Sigma^2}{\sigma^2} \quad (20)$$

The main result of our theory, Eq.19 is noteworthy. First, it shows that the number of mature B cells decays exponentially with respect to the maturation threshold $G_c$. This immediately implies that there is a ceiling in affinity of matured BCRs achieved when $N_{AM}^{2s} \leq 1$, Such ceiling was indeed observed in experiments (Batista and Neuberger 1998). Second, the theory predicts that greater diversity of Ab interactions in the germline pool $(\Sigma)$ is advantageous for selection of potent antibodies. However this consideration should be balanced against possible detrimental effect of promiscuity of Ab interactions including autoimmunity effect (Sun et al. 2005). The analysis of the balance between affinity maturation and autoimmunity is a subject for future study.

Finally, our theory provides insight into the fate of B cells in response to incoming antigens. The affinity of most incoming Ags to germline Immunoglobulins is close to $G_{av}$ (which depends on incoming antigen). Our result Eq.20 shows that SHM is efficient and necessary only when germline Igs exhibit relatively low affinity to the incoming Ag. (i.e. when $G_{av} - G_c$ is large enough.). In the opposite case when incoming Ags already have substantial affinity to germline Igs, differentiation and direct clonal expansion of potent B cells in germline pool is advantageous. Remarkably, it appears that Nature follows similar considerations. (Hangartner et al. 2006).

A more quantitative analysis of the results of presented theory requires a systematic study of sequence diversity and energetics of interactions between germline and mature Igs and various Ags (Tomlinson et al. 1996; Chen et al. 1999; Clark et al. 2006) which will be addressed in future studies.

**Controls: What is important for effective immune response?**

Our model of immune response is minimalistic yet it contains a number of assumptions such as Ag and ''self-interaction'' (autoimmunity) dependent activation of B-cells. Importantly and in contrast with earlier models (e.g. (Pierre et al. 1997)) our model explicitly considers proteins stability. By requiring that potent Ig and viral proteins must be stable in their native conformations our model introduces competition and interrelation between folding and binding. In order to understand the effect of each of those requirements we ran control simulations whereby individual assumptions of the model were relaxed one by one and their impacts on various aspects of immune response in the model were examined. (Table I). Apparently activation is crucial for effective immune response – removal of the dependence of duplication rate of B cells on Ag affinity results in deteriorated immune response. In contrast selection against autoimmunity imposes a strict constraint on efficiency of immune response by dramatically decreasing the probability of the healing outcome.

Our simulations revealed a specific mechanism of evolution of potent Igs in response to mutating viral Ag. In particular we found that mature Igs are progenitors of a small subset of naïve cells which exhibited initial affinity to the incoming Ag higher than certain threshold (''Monoclonality'') despite the fact that sequence diversity of matured Ig's is significant (see Fig.6), in agreement with experiment (Jacob et al. 1991). Interestingly, we find that the physical reason for monoclonality of evolved potent Igs in simulations is in the interplay between Ig stability and its affinity to the antigen. Indeed when the requirement of Ig stability is relaxed in control simulations the mechanism changed dramatically – ''monoclonality'' is lost, potent Igs are now ancestors of a broad variety of germline Igs and activation times (the number of time steps to achieve mature Igs) drop dramatically. We conclude that interplay between folding and binding shapes the search in sequence space of Igs in a crucial way. Consequently, relaxation of requirement for Ig proteins to be stable leads to much greater polyclonality and sequence diversity of mature Igs.

|  | Full Model |  | Immunoglobulin |  |  |  |  |  | Virus |  |
|---|---|---|---|---|---|---|---|---|---|---|
|  |  |  | w/o stability |  | w/o activation |  | w/o autoimmunity |  | w/o stability |  |
|  | H | P | H | P | H | P | H | P | H | P |
| Percentage probability | 48.5 | 21.0 | 52.0 | 48.0 | 1.0 | 1.0 | 88.5 | 11.5 | 23.5 | 49.0 |
| Monoclonality | 0.79 | 0.78 | 0.16 | 0.17 | 0.06 | 0.04 | 0.80 | 0.83 | 0.77 | 0.86 |
| $<P_{int}(Ig,v)>$ (stdev) | 0.54 (0.10) | 0.51 (0.12) | 0.41 (0.18) | 0.40 (0.18) | 0.17 (0.07) | 0.15 (0.05) | 0.55 (0.11) | 0.60 (0.13) | 0.54 (0.10) | 0.54 (0.11) |
| Sequence diversity | 0.49 | 0.43 | 0.70 | 0.70 | 0.47 | 0.45 | 0.77 | 0.61 | 0.55 | 0.44 |
| Activation time | 173 | 184 | 55 | 52 | 36 | 20 | 268 | 204 | 234 | 177 |

**Table 1**. *Control simulations where various constraints of the immunity model have been relaxed. 200 independent simulations are performed at $m_v=0.0003$ and $m_{Ig}=0.08$ for each control case. H and P respectively represent healing and persistent infection pathways. Monoclonality (or monoclonal selectivity) is defined as the maximum fraction of the activated "red antibodies" which are averaged over all pathways of healing or persistent infection. $<P_{int}(Ig,v)>$ is the averaged protein-protein interaction strength between Ig and virus protein, and the numbers in parenthesis are standard deviations of $P_{int}(Ig,v)$. The sequence diversity is averaged sequence entropy of antibodies when the activated "red antibodies" first achieve the maximum fraction of the antibody pool of system*

*during simulation. The activation time is the elapsed time for the activated "red antibodies" to achieve the maximum fraction pool from the first viral infection.*

**Conclusions**
We presented a microscopic bottom-up model of immunity in an interacting host-pathogen system. A complex interplay between genetics, physics of proteins, and population dynamics provides a rich variety of dynamical regimes, identifying all possible outcomes such as healing, host extinction, and persistent infection as a function of the mutation and replication rates. The presented diagram of the infection outcomes as function of the mutation rates can be of interest for the development of therapeutic approaches to certain diseases.

Perhaps the most striking result of our study is that two stage immune response of clonal selection and affinity maturation exactly as postulated by Burnet has emerged from simulations in a purely microscopic model which did not postulate a'priori any mechanism. Importantly, we showed that the Burnet's mechanism is a consequence of delicate balance between search and selection of high affinity antibodies and the need to maintain their stability as folded proteins. As a result sequence evolution of Ig proteins included initial selection of moderately binding germline molecules and their subsequent clonal amplification accompanied by affinity maturation (Jacob et al. 1991). As our control shows, neglect of stability factor dramatically changes the dynamics of affinity maturation leading to highly polyclonal Ig molecules (our model corresponds to development of just one GC). However, we stress here that monoclonality of emerged Ig molecules does not result in their sequence homogeneity – significant sequence variation of matured Ig molecules is observed both in our model and in reality (MacLennan 1994).

The analysis of affinity maturation in simulations motivated us to develop an analytical theory of affinity maturation. Its key insight is a non-trivial quantitative relation between molecular properties of individual V domain genes in centroblasts - such as the number of acquired mutations (Jacob et al. 1991) and their energetic impact– and the number of generations of centroblast expansion in GC required to achieve successful AM. Our results on dynamics of affinity maturation are in excellent quantitative agreement with experimental observations, despite simplicity of the model and theory. In our simulations Ig molecules accrue in average 5 mutations during their affinity maturation process while in reality it is closer to $mt$=10-15 mutations (Shih et al. 2002). This can be expected because our model proteins are smaller than V domains of Ig. We assume that SHM program operates close to optimal. The estimate $h = 1 kcal/m$ and $\sigma^2 = 3(kcal/m)^2$ for the parameters of the model comes from the analysis on impact of point mutations on protein stability (Zeldovich et al. 2007a). Eq. 13 then predicts that upon completion of the SHM program the fraction of mature B-cells in the pool is $P_{AM} = 10^{-4} \div 10^{-5}$ (the variation is due to uncertainty in the experimental estimates of the number of mutations in $V_{H,L}$ (Shih et al. 2002) (Jacob et al. 1991)). Therefore our theory predicts that B cells should expand in GC to population of $10^4 \div 10^5$ in order to ensure successful AM. Again this number is entirely consistent with reality: cell cycle of centroblasts is about 6 hrs (MacLennan 1994), though a longer time closer to 12hrs has been suggested recently (Allen et al. 2007), and the SHM rate is roughly 1 mutation per V domain per generation (Kleinstein et al. 2003). Observed number of mutations is therefore accrued in 15 generations (4-6 days) - exactly the time frame to generate $10^4 \div 10^5$ B cells (MacLennan 1994; Or-Guil et al. 2007). This analysis shows that AM is likely to be a ''one-pass'' process in each GC. Perelson and coworkers argued that a ''one-pass'' AM is hardly efficient (Oprea et al. 2000). However their mathematical model is based on a number of assumptions (e.g two affinity classes, no post-SHM expansion in LZ). In contrast, here we consider a continuum of mutation effects in a physically realistic model of protein-protein interactions, not just two affinity classes. Further, we note that AM is successful even when few B cells expressing sufficiently potent Igs appear in the pool after SHM. They can differentiate into plasma cells and expand (Phan et al. 2006). Our analysis shows that an optimal SHM program depends dramatically on the initial affinity of Igs at the start of the program (parameter $G_0$, see Eq.4). This means that for multiple round SHM to be efficient, the SHM program should adjust at every round as Igs of higher affinity progressively evolve. While not impossible, there is no experimental evidence that such adjustments take place. Without them, however, the progressive rounds of SHM of fixed length may quickly become detrimental as they would subject evolving potent Igs to multiple rounds of random destabilizing mutations.

(Shih et al. 2002) argued that SHM program is fixed in an organism. Our theory predicts that optimal program – the average number of attempted mutations – depends on two affinity thresholds for T-dependent B cell activation – the first (at affinity $G_0$) is when resting B cells are activated to enter the response, and the second ($G_c$) which selects those somatically mutated Immunoglobulines which acquired sufficient affinity to the Ag for differentiation into plasma and/or memory cells. It is likely that each organism establishes these thresholds so that it can run an optimal SHM program. However a fraction of germline B cells may express potent antibodies with binding energy to incoming Ag below threshold value $G_0$. In this case it could be detrimental for these B cells to enter SHM program. Rather these B cells undergo extrafollicular plasma cell differentiation without SHM (Paus et al. 2006; Phan et al. 2006). An interesting recent observation suggests that indeed SHM may be individually tuned in each organism and that such tuning depends on relationship between affinity of germline Igs and mature Igs as predicted by this theory. It was shown in (Dooley et al. 2006) that in ectothermic vertebrates (sharks) the initial affinity of IgNAR is very high – $K_D$ in low nm; however affinity maturation is short (sharks lack Germinal Centers) and improves affinity only an order of magnitude, while in vertebrates initial affinity is weak – in millimolar to micromolar range - but affinity maturation improves $K_D$ 1000 fold or more (De Genst et al. 2004; Dooley et al. 2006). Shorter affinity maturation in shark compared to vertebrates is related to slower rate of their B cell replication, however it is compensated by high rate of the initial IgNAR mutation to create an extremely diverse initial germline pool that contained highly potent antibodies. Our theory predicts exactly that – that diversity of Ag-Ig interactions in the germline pool (parameter $\Sigma$ in Eqs(17-20)) is a major determinant of the duration and depth of the SHM program (e.g. see Eq.20).

Our simulations show, paradoxically, that cytopathic viruses evolve to have mild affinity to germline Igs. Indeed as shown in Figure 5 mutating viruses acquire selective advantage by developing moderate binding to germline Igs of the organism (in simulations shown in Fig.5 we kept Ig mutation rate at 0). This compromise helps coexistence as neither party eliminates the other one. Interestingly a similar behavior is observed for natural cytopathic viruses (e.g VSV in mice) – in this case the defense is realized mostly due to significant affinity of viral antigens to germline Igs of the host organism without affinity maturation (Hangartner et al. 2006). What about SHM response in this case when incoming antigens have strong affinity in average to most germline BCR? The theory predicts that in this case (high $G_{av}$ in Eq.18) the ''one-shot'' selection where potent Igs are directly selected from the germline pool is a winning strategy. Experiment shows that this is indeed the case: infection with highly cytopathic VSV virus induces direct response from the germline without much affinity maturation while infection with poorly cytopathic viruses (LCMV in mice, HCV,HBV in humans) induces a response which includes affinity maturation via SHM (Kalinke et al. 1996; Hangartner et al. 2006).

An important aspect of our microscopic model is consideration of autoimmunity which significantly restricts possible scenarios of sequence evolution of Ig. While real organisms defend themselves against autoimmunity through a complex process of clonal elimination of helper T-cells in the thymus (Kappler et al. 1987), here we model it in a coarse-grained way by postulating that B-cell activation is restricted when Immunoglobulins which they produce attack the organism's own antigens. While the process of T-cell maturation is not yet fully understood, this coarse-grained description captures its essence without introducing T-helpers explicitly in the model. Deem and coworkers suggested that cross-reactivity significantly affects affinity maturation dynamics (Sun et al. 2005). Our simulations are in agreement with this proposal. Indeed as can be seen in Table I autoimmunity significantly constrains the sequence repertoire of emerging Igs and removing autoimmunity requirements shifts the likely infection outcome from persistent infection to healing even for highly mutating viruses.

Quantitatively, our model suggests that in the healing regime the optimal mutation rate of an antibody should be about $10^3$ higher than that of a virus. This finding is generally in agreement with experiment: the SHM rate is about $10^{-3}$ per bp per division (Kleinstein et al. 2003), while DNA viruses have a much lower mutation rate, $10^{-6}$ to $10^{-8}$ per bp per replication, at least three orders of magnitude lower (Drake et al. 1998). RNA viruses, however, have much higher mutation rates (Drake and Holland 1999), and, according to the model, may often cause a persistent infection. Quite remarkably the optimal Ig mutation rate observed here almost exactly coincides with estimates for SHM in natural Immunoglobulins (see Fig.S3 in Supporting Information).

This model provides a unique opportunity to glean insights into evolution of adaptive immunity by highlighting bare bone minimal requirements for a functional adaptive immune response. Our assumptions here are simple: an interaction between organism's defensive proteins (Igs) and viral antigen 1) imparts viral ability to grow and 2) affects the rate at which Ig producing B cells divide (activation). While the first assumption is mechanistic and almost trivial, the second one, being crucial for the model to work (see Table I), is not so ''evolutionary innocent''. Indeed the specific molecular mechanism by which affinity to antigen affects the fate of B cells, especially their apoptosis and T-cell dependent expansion is elusive (Batista and Neuberger 1998; McHeyzer-Williams et al. 2006). Our results suggest that evolutionary discovery of such feedback mechanism had been the key punctuated step towards adaptive immunity. It would be extremely interesting to get further insights from phylogenomics of how such mechanism emerged and what are its precursors in invertebrates.

Earlier theoretical studies of many aspects of immune response, including affinity maturation, treated protein thermodynamics and interactions within simplified phenomenological frameworks. A ''string approximation'' which does not consider protein stability and treats interacting surfaces as superimposed short strings of simplified amino acids – ''characters'' with simple ''matching rules'' to approximate interaction energetics was used in (Celada and Seiden 1996; Pierre et al. 1997). Another widely used approximation is Kauffman's NK model which presents various contributions to protein stability and interactions as independent random quantities (Kauffman and Weinberger 1989). Although these models are useful as first approximations, their generic limitation is that the properties of proteins are not derived from their sequences. However evolution of potent antibodies amounts to improving their affinity to an antigen while keeping them from attacking the organism's own proteins (autoimmunity) and retaining their stable structure. This delicate balancing act has to be accomplished via sequence search. Therefore the basic principles of adaptive immunity can be consistently addressed only in sequence-based models which explicitly derive Ig stability and interactions from their sequences using a physically realistic energetics of intra- and intermolecular interactions. The physical realism of energetics is an important issue. While string matching models assume linear arrangement of amino acids on the interface and that each amino acid should have a unique partner in a potent interface, in reality interaction energy is a sum of many relatively weak contributions, which can be realized, for each amino acid, by diverse and multiple interaction partners (Janin and Seraphin 2003; Bahadur et al. 2004; Bordner and Abagyan 2005; Lukatsky et al. 2007). This fundamental property of interactions in proteins gives rise to significant plasticity in sequence space which crucially affects sequence evolution of Igs via development of neutral networks (Bornberg-Bauer and Chan 1999; Zeldovich et al. 2007b). While our model is coarse-grained, it certainly captures these key aspects of proteins, providing insights into evolution and selection of potent Ig sequences. Most importantly, the combination of ab initio simulation with analytical theory helps to get better understanding of robust principles of adaptive immunity. In particular we found that SHM is a sensitive mechanism which may be effective only under certain well-defined conditions, such as affinity range of incoming antigens (Hangartner et al. 2006; Paus et al. 2006). There is significant evidence that Nature indeed employs SHM selectively only when conditions are conducive to productive SHM program. The molecular mechanism which helps B cells to react selectively to incoming antigens is not clear yet.

Our model is still quite minimalistic. Its particular molecular mechanisms here are quite schematic and may differ significantly from actual biological pathways that are operational in living cells. In particular we do not explicitly model the important role of helper T-cells in B-cell activation. Rather we assume this mechanism in a coarse-grained fashion by postulating a certain relationship between B-cell replication rate and affinity of their Ig to an antigen. Further the spatial dynamics aspects are completely disregarded here – perhaps this factor leads to exaggerated effects in populations of viruses and cells showing a dramatic drop in population sizes at the healing bottleneck (Fig.3). However this aspect of the present model is less realistic as it does not treat tissues as real 3D objects where proliferation of infection is limited by diffusion. Another limitation of this study is lack of consideration of dosage and memory effects. This can be done in the present model by introducing concentrations of antigens and Igs and using the law of mass action to estimate binding, although this refinement goes beyond the scope of this paper (work in progress). Despite these limitations, the striking qualitative and quantitative similarity of the general features of immune response found in this model to natural defense mechanisms attest to the robustness of physical principles of adaptive immunity. Quite complicated machinery evolved to realize these physical principles in jawed vertebrates. Our simulations first and foremost show an amazing power of evolution where a blueprint for a very complex function is provided by physical principles while its execution is brilliantly carried out via mutation and natural selection processes as postulated by Darwin long ago.

**Model and Methods**

The model (see Figure 1, and Supplementary Information for details) consists of three types of entities, each with its own population dynamics: host cells, Immunoglobulins (antibodies) and viruses. Each host cell, antibody, and virus are modeled individually by keeping track of their protein sequences. Proteins are modeled as 27-unit compact polymers, so that their thermodynamic properties, stability and interaction can be calculated exactly from their primary sequences (More details are in Supplementary Information and in (Zeldovich et al. 2007b)). A genome of each host cell consists of three genes. The replication rate of cells is constant, and the (intrinsic) death rate of host cells is determined by stability of its proteins as in (Zeldovich et al. 2007b). Each virus carries one protein (antigen); once a virus infects a host cell, the virus starts to replicate at a rate, $b^v$, dependent on the stability of viral protein and the interaction strength between viral and Ig proteins.

$$b^v = b_0^v \cdot P_{nat}(v) \cdot \left(1 - P_{int}(Ig,v)\right), \tag{21}$$

where $b_0^v$ is a viral replication rate constant, $P_{nat}(v)$ is protein stability of virus (i.e. Boltzmann probability for the virus protein to be in its native conformation), and $P_{int}(Ig,v)$ is Boltzmann probability of a specific interaction between Ig protein and virus protein which serves as a measure of their interaction strength (see Supplementary Text for exact definition of this quantity). Strong binding inhibits virus replication. We model cytophatic visruses which may destroy host cells via lysis. Once the number of viruses in a given cell exceeds the lysis threshold, the cell dies and releases the viruses that can further infect uninfected cells. If free viral particles cannot find any host cell within a certain time, they are removed. The replication rate of antibodies $b^{Ig}$ is determined by interaction between Ig and viral proteins if the cell is infected (strong binding increases antibody replication rate), and the autoimmunity effect (strong binding between Ig and a functional protein decreases the replication rate),

$$b^{Ig} = b_0^{Ig}[1 + \alpha P_{int}(Ig,v)] \cdot \left[1 - \max_i P_{int}(Ig,g_i)\right], \tag{22}$$

where $P_{int}(Ig,g_i)$ and $P_{int}(Ig,v)$ are respectively the Ig protein's interaction strength with $i$-th protein of the cell and a randomly chosen viral particle in the cell. The index $i$ runs from 1 to 3. In the context of the immune system, an increase of antibody replication rate with the antigen binding strength is a well-known phenomenon of B-cell activation by an antigen (MacLennan 1994; Rajewsky 1996; McHeyzer-Williams et al. 2006).

Upon division of an infected cell, viruses are randomly distributed between its two off-springs, and when an infected cell dies, any viruses it contains are also removed from the population. The simulation ensemble consists of up to 5000 completely independent cells, 20000 independent virus particles, and 8000 antibodies, mutating and replicating according to the above rules. If necessary, cell population is clipped at 5000, simulating a chemostat, and the antibody population is clipped at 8000. The mutation rate of Ig proteins is set significantly above that of functional proteins, mimicking SHM observed in B-cells, and the mutation rate of viruses is also higher than the normal host protein mutation rate. For the initial 1500 evolutionary time steps, host cell genes and antibodies are allowed to evolve with mutation rates $m_{cell}=m_{Ig}=0.05$ per protein per time step to equilibrate their protein sequences, and then the mutation rates drop to $m_{cell}=0.0001$ and $m_{Ig}=0.005$ (see Supplementary Information for details). At $t=2001$, one thousand of identical free viruses are introduced into the system, starting the infection. In order to evaluate how mutation rates of Ig genes and viruses affect the outcome of cell-virus competition, we run multiple simulations with different mutation rates of viruses and Ig genes while keeping mutation rates of host cell's genes fixed as described above.

*Acknowledgememts.* This work is supported by the NIH. We thank Michael Deem for valuable comments on the manuscript.

## Supplementary Text

## Methods

**Protein model: Stability and interactions**

We consider a 3x3x3 lattice protein as a model protein in order to calculate thermodynamic quantities such as protein stability and protein interaction strength exactly (Shakhnovich 1990; Zeldovich et al. 2007). To save computer time, we reduce protein structural space from 103,346 structures to randomly and uniformly chosen 10,000 structures. Each protein folds into the native structure which has minimum energy out of 10,000 structures. Protein stability is calculated in terms of the Boltzmann probability ($P_{nat}$) of the native (i.e. lowest energy) structure of the protein at given temperature,

$$P_{nat}(g,T) = \frac{\exp[-E_0(g)/T]}{\sum_{i=1}^{10000} \exp[-E_i(g)/T]}, \quad (S1)$$

where $g$ is a protein sequence in 20 amino acid type alphabet, $E_0(g)$ and $E_i(g)$ are energies of protein in the native and $i$-th structure respectively, and $T$ is an environmental temperature. In order to calculate interaction strength between two proteins, we consider rigid-body docking between two 3x3x3 lattice proteins. Each lattice protein has six binding surfaces and four-fold rotational symmetry of a binding surface (only the strongest binding positions when two faces overlap completely making 9 interaction bonds are taken into account. (see Figure S1) Therefore, a pair of lattice proteins has 6x6x4 binding conformations. Protein interaction strength is defined by the Boltzmann probability ($P_{int}$) of the specific binding between two protein $g_1$ and $g_2$ as following.

$$P_{int}(g_1,g_2) = \frac{\exp[-f \cdot E_0(g_1,g_2)/T]}{\sum_{j=1}^{144} \exp[-f \cdot E_j(g_1,g_2)/T]}, \quad (S2)$$

where $E_0(g_1, g_2)$ is the binding energy of two proteins in the binding mode corresponding to lowest energy out of all 144 possible binding modes, and $f$ is a pre-factor that takes into account possible different strengths of intra-protein and inter-protein interactions. We use the Miyazawa-Jernigan pairwise contact potential for both protein structural and interaction energies (Miyazawa and Jernigan 1996), but scale protein-protein interactions by a constant factor. We chose $T = 0.85$ in Miyazawa-Jernigan potential dimensionless energy units and set $f = 1.5$.

**Cell dynamics; reproduction, death**

We use here a simple model of host cells cell whose population dynamics is described by genetic information encoded in its genome. The genome of a cell consists of three genes, whose protein stability governs death rate the host cells. We connect the minimum stability of any protein in a cell to its death rate $d$ as follows (Zeldovich et al. 2007):

$$d = d_0 \left[ 1 - \min_i P_{nat}(g_i) \right], \tag{S3}$$

where $d_0$ is a cell death rate constant and $P_{nat}(g_i)$ is $g_i$ protein's stability and index $i$ runs from 1 to 3. These cells also replicate at a constant rate.

**The immune system; immunoglobulin (Ig) dynamics**

For simplicity, we coarse-grain the complex processes of antigen recognition, B-cell activation, and antibody production into the production rate of Ig proteins, determined by their physical interactions with the antigen (virus) and host cell proteins. The interaction between Ig protein and viral protein is responsible for the molecular recognition and suppression of the spread of exogenous toxic materials. Once an organism is infected by a virus, it can obtain immunity against the virus by increasing the strength of interaction between Ig and viral RNA or enzymes. Also, the Ig protein should recognize "self" genes from "non-self" genes. If it mistakes "self" proteins for "non-self" proteins and binds them strongly, it suppresses the function of cellular proteins. This is the main cause of autoimmunity, and Ig's evolution to avoid autoimmune disease. We mimic this by incorporating the protein interaction strength between Ig and normal proteins of the cell to the replication rate of Ig, which is written as follows.

$$b^{Ig} = b_0^{Ig} \cdot \left[ 1 - \max_i P_{int}(Ig, g_i) \right], \tag{S4}$$

where $b_0^{Ig}$ is an Ig protein's replication rate constant and $P_{int}(Ig, g_i)$ is the interaction strength between Ig and $i$-th normal protein in the cell and index $i$ runs from 1 to 3. The presence of antigens activates the antigen-specific B-cell's differentiation into plasma cells to elevate the load of antibodies matching the antigens. The amplification of the antigen-specific B-cell differentiation is described by effective activation which results in enhanced the replication rate of B cells that produce antibodies with high specificity to the antigens. The infection by virus modifies the replication rate of antibody-producing B-cells (and therefore the rate of effective ''replication'' of Ig proteins) as follows:

$$b^{Ig} = b_0^{Ig} \cdot \left[ 1 - \max_i P_{int}(Ig, g_i) \right] \cdot \left[ 1 + \alpha \cdot P_{int}(Ig, v) \right], \tag{S5}$$

where $P_{int}(Ig, v)$ is Ig protein's interaction strength with viral particle in the system and $\alpha$ is antibody activation factor by antigens. Ig proteins suppress viral replication only when they reside in infected cells. Ig proteins can be either bound to a viral protein within a host cell, or remain free. The replication rate of the free Ig proteins is determined by calculating their interaction strength with a randomly chosen viral protein for $P_{int}(Ig,v)$ and also randomly chosen cell for max $P_{int}(Ig,g_i)$. Free Ig proteins have no direct effect on the population dynamics of host cells and viruses; at each simulation step, a free Ig can penetrate an infected host cell and bind to one of the infecting viruses. Bound Ig cannot escape from the cells; upon division of an infected host cell, viruses together with any bound Ig are randomly distributed between the daughter cells. When an infected host cell dies, any Ig proteins it contained are removed. An Ig protein can fold and has functionality when its stability $P_{nat}$ (see Eq.S1) is greater than 0.4 in the model. Mutations which decrease Ig stability $P_{nat}$ below 0.4 lead to the "death" (removal) of the corresponding antibody.

**Virus dynamics, replication and lysis**

A viral particle carries a single protein (antigen), denoted by *v*. A free viral particle can infect a cell which is unoccupied by other viral particles with viral infection rate. This particle can be mutated or replicated only when it resides in an infected host cell. Free viral particles are quickly removed when they fail to find their host cells. When the number of viral particles in a living host cell exceeds the lysis threshold, the viral particles induce lysis of the cell and are released. The replication rate of the virus is regulated by its protein stability and protein interaction strength with Ig protein. A viral replication rate $b^v$ can be written as following.

$$b^v = b_0^v \cdot P_{nat}(v) \cdot \left[1 - P_{int}(Ig,v)\right], \tag{S6}$$

where $b_0^v$ is a viral replication rate constant, $P_{nat}(v)$ is stability of the viral protein, and $P_{int}(Ig,v)$ is protein interaction strength between Ig and viral protein.

**Simulations**

Initially 100 cells with 3 identical primordial genes and 100 identical Ig proteins are seeded in the system. Initial simulation starts with same cell division rate *b*, cell death rate *d*, and Ig protein's replication rate $b^{Ig}$ ($b=d=b^{Ig}=0.1$) and the system evolves to achieve high stabilities for normal proteins of cells and their weak interaction with Ig proteins to minimize the autoimmunity effects. Due to the differences between the time scales of protein evolution and a much faster response to a viral infection, the population dynamic simulation proceeds in two steps. For 1500 initial time steps, the system including cells and antibodies are allowed to evolve with mutation rate of functional proteins of the cell and Ig proteins, $m_{cell}=m_{Ig}=0.05$ (*attempted* mutations per gene per unit time) in order to equilibrate the system. At $t=1501$, we select one cell whose averaged maximum $P_{int}(Ig,g_i)$ over all antibodies in the system has lowest value and replicate it 5000 times in order for all cells to have the same genome at the moment of virus infection. All other cells are removed. Then, the *attempted* mutation rates of functional proteins of the cell and Ig proteins is reduced to $m_{cell}=0.0001$ and $m_{Ig}=0.005$ respectively, and the system evolves for 500 time steps to re-equilibrate at new mutation rates. At $t=2001$, 1000 identical free viruses are introduced in the system, starting the infection with the viral infection rate, 0.75. The sequence of viral protein is designed for the initial stability $P_{nat}(v)=0.66$ and the averaged initial protein interaction strength between Ig proteins and viral protein is $P_{int}(Ig,v)=0.15$. The viral replication rate constant $b_0^v$ and the lysis threshold are set to 0.79 and 4 respectively. After infection, the system evolves up to $t=4000$. In reality, B-cells divide 3~4-fold faster when activated than normal (Janeway et al. 2001) and we set B-cell activation factor to $α=5.0$ to obtain 3~4-fold faster activation at a moderate antibody-antigen interactions strengths $P_{int}(Ig,v)=0.4~0.7$.

**Sequence entropy calculation**

In order to analyze the degree of diversity of Ig proteins, we calculated the sequence entropy of Ig proteins. The sequence entropy of a residue in *k*-position is defined as following (Yano and Hasegawa 1974).

$$S_k = -\sum_{i=1}^{20} P_i^k \log P_i^k \tag{S7}$$

, where $P_i^k$ is frequency of amino acid of type i in *k*-position in a multiple sequence alignment. The sequence entropy for the whole Ig protein is obtained by averaging over all 27 positions in its sequence.

## References for the Supplementary Text

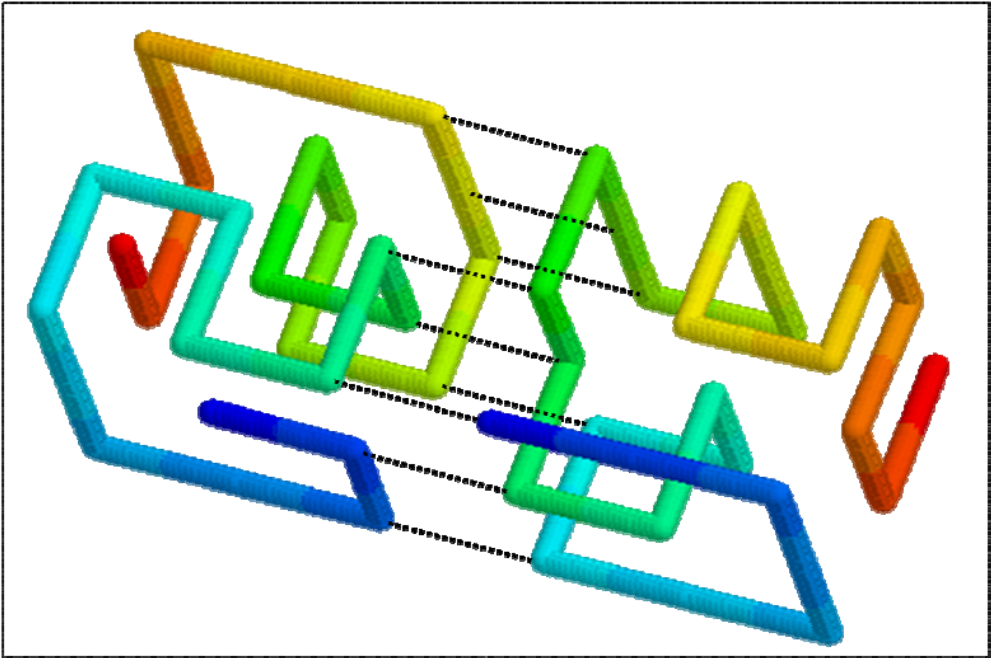

Figure S1.
**The snapshot of two interacting lattice proteins.** Two proteins in 3x3x3 lattice are presented in backbone diagram. Two proteins are interacting with each other as sharing 9 interaction bonds, which is presented with black dotted lines.

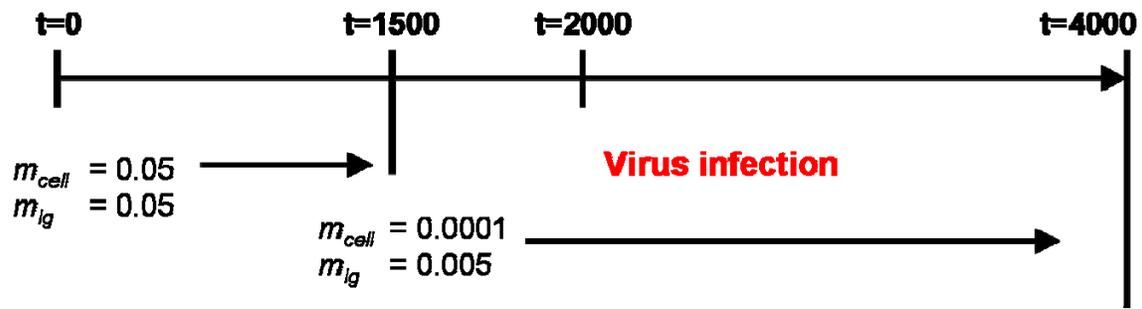

Figure S2.
**The simulation procedure.** The initial simulation runs up to t=1500 with the high mutation rates of cells and antibodies ($m_{cell}=m_{Ig}=0.05$ *attempted* mutations per gene per unit time). And then, the *attempted* mutation rates of functional proteins of the cell and Ig proteins is reduced to $m_{cell}=0.0001$ and $m_{Ig}=0.005$ respectively, and the system evolves for 500 time steps to re-equilibrate it. At t=2001, 1000 identical virus particles starts to infect cells in the system.

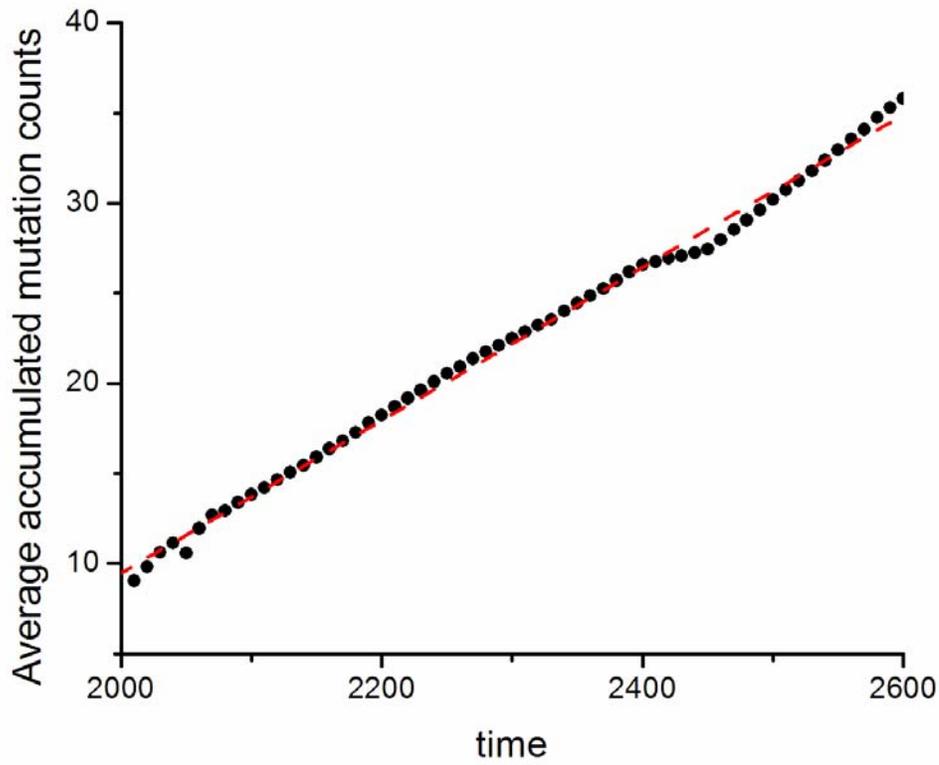

Figure S3.

**Observed mutation rate of Ig gene.** The black dots represent the average accumulated number of mutations in Ig genes in the healing case ($m_v$=0.0003 and $m_{Ig}$=0.08). A linear fit estimate (red dashed line) corresponds to the observed mutation rate of Ig gene $\tilde{m}_{Ig}$=0.04 (mutations per gene of 81 nucleotides per time step). In the presence of virus, the division rate of (activated) B-cells increases to ~0.4, so B-cell division time is 2.5 time steps. Therefore, the observed mutation rate of Ig gene is at least $1.2 \times 10^{-3}$ mutations per bp per division (neglecting the redundancy of the genetic code), which is comparable to the somatic hypermutation rates estimated by Kleinstein